\documentclass[aip,jcp,numerical,reprint,floatfix]{revtex4-1}
\usepackage[utf8]{inputenc}
\usepackage{graphicx}
\usepackage{amsmath}
\usepackage{color}
\usepackage{array}

\begin{document}

\title{Close packing of rods on spherical surfaces}
\author{Frank Smallenburg}
\author{Hartmut L\"owen}
\affiliation{Institut f\"ur Theoretische Physik II: Weiche Materie, Heinrich-Heine-Universit\"at D\"usseldorf, Universit\"atsstr. 1,
40225 D\"usseldorf, Germany}

\begin{abstract}
We study the optimal packing of short, hard spherocylinders confined to lie tangential to a spherical surface, using simulated annealing and molecular dynamics simulations. For clusters of up to twelve particles, we map out the changes in the geometry of the closest-packed configuration as a function of the aspect ratio $L/D$, where $L$ is the cylinder length and $D$ the diameter of the rods. We find a rich variety of cluster structures. For larger clusters, we find that the best-packed configurations up to around 100 particles are highly dependent on the exact number of particles and aspect ratio. For even larger clusters, we find largely disordered clusters for very short rods ($L/D = 0.25$), while slightly longer rods ($L/D = 0.5$ or $1$) prefer a global baseball-like geometry of smectic-like domains, similar to the behavior of large-scale nematic shells. Our results provide predictions for experimentally realizable systems of colloidal rods trapped at the interface of emulsion droplets.
\end{abstract}

\maketitle

\section{Introduction}

It is a well-known fact that the optimal way of packing spherical particles on a flat substrate is a hexagonal lattice. However, when the substrate is curved, the interplay between surface curvature and particle packing leads to frustration and lattice defects \cite{nelson2002defects, bowick2009two, irvine2012fractionalization,bausch2003grain,mughal2012dense,irvine2010pleats}. For example, placing spheres on a spherical substrate leads to a minimum of 12 point defects in the hexagonal lattice structure, and higher numbers of defects typically occur in optimally packed structures~\cite{bausch2003grain}. As such, the optimal packing of a given number of spheres on a spherical surface is far from trivial to predict.

Although the packing of objects on spherical surfaces has received considerable attention as a mathematical problem~\cite{toth2013lagerungen,delsarte1977spherical,kottwitz1991densest,Sloane}, it has recently also drawn attention in the field of colloidal self-assembly, due to the experimental realization of colloidosomes: spherical shells of colloidal particles self-assembled on the surface of an emulsion droplet~\cite{rossier2009colloidosomes}. In these systems, colloids are held at the oil-water interface by the drive of the system to minimize the oil-water interfacial free energy~\cite{manoharan2006colloidal}, or trapped in thin shell-like droplets occuring in a double emulsion~\cite{lee2008double,PhysRevLett.99.157801,zhou2013thermally}. In both scenarios, the interfacial energies associated with the interfaces between oil, water, and particles are typically much higher than the thermal energy fluctuations in the system, and as a result the motion of the particles is effectively confined to a rigid spherical shell, resulting in a direct realization of the spherical packing problem. Dense surface packings in such systems can be achieved by slowly evaporating the discontinuous phase in the emulsion, resulting in shrinking droplets which compress the colloids into a compact cluster~\cite{manoharan2003dense}, or simply by waiting for sufficient numbers of colloidal particles to attach to the interface~\cite{dinsmore2002colloidosomes,lee2008double}. When many spherical colloids are attached to a single droplet, the resulting structure is a hollow spherical shell of particles whose packing includes the defect structures expected in spherical packings. Typical suggested applications for these shells are aimed at food science and drug delivery \cite{dinsmore2002colloidosomes,zhou2013thermally}, with the aim of selectively containing or releasing the contents of the shell, or to enhance emulsification. As a result, a key property of these colloidosomes is the fraction of the droplet surface covered by particles, and the size of the pores between the particles.  On the other hand, when only a few colloids are attached to a droplet, they form a sequence of polyhedral clusters of varying symmetry~\cite{manoharan2003dense,cho2005colloidal,Sloane}. It has been suggested that such small colloidal clusters can in principle be used as building blocks for further self-assembly processes, potentially opening the way to new self-assembled structures~\cite{zerrouki2006preparation}. For this purpose, the structure of the clusters is key in controlling the properties of the final material.

The structure and surface coverage of a spherical packing can be drastically altered by changing the shape of the particles used.  In particular, replacing the spheres with rod-like particles opens up the possibility of liquid crystalline order in the resulting spherical shell. Recent experiments have demonstrated several ways of generating thin nematic or smectic shells~\cite{lopez2011drops}, including the attachment of particles to the interface of  air bubbles~\cite{zhou2009rigid,may2012dynamics}, as well as double emulsions where rod-like particles are confined to the spherical shell between two concentric droplets~\cite{lopez2011frustrated,zhou2013thermally,PhysRevLett.99.157801, lopez2012smectic,liang2013tuning,liang2011nematic,lopez2012smectic,koning2013bivalent}. 

\begin{figure}
  \includegraphics[width=0.95\linewidth]{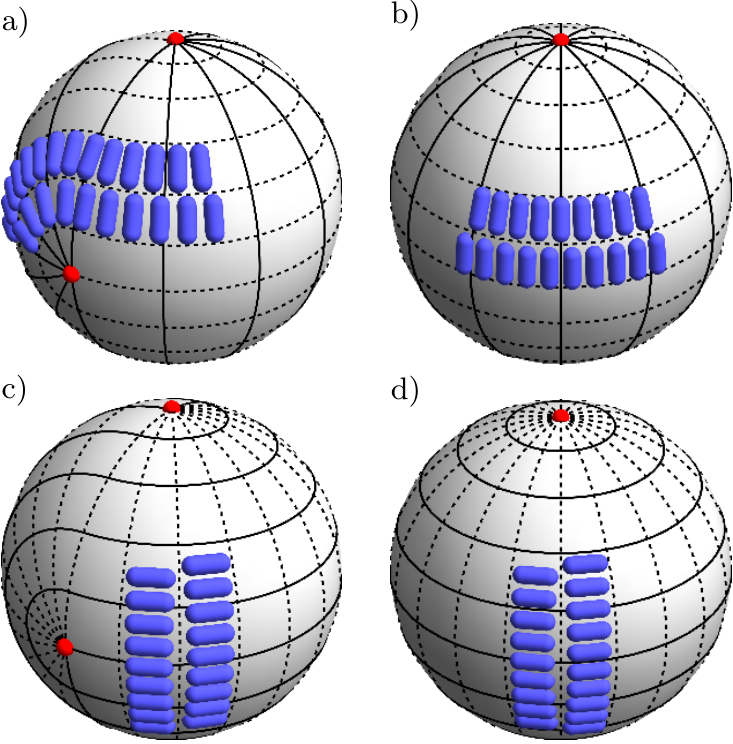}
 \caption{Orientation fields for rods on a spherical surface, with a few rods drawn in to illustrate their arrangement. The solid lines indicate the orientation of the particles, and the dashed lines indicate the direction of  layers of particles (as would occur in a smectic or crystalline domain). The topological defects are indicated in red. The fields correspond to {\bf a)} baseball geometry, {\bf b)} polar geometry, {\bf c)} antibaseball geometry, and {\bf d)} antipolar geometry. Note that in (a) and (c), the four defects are placed on a great circle. }\label{fig:directorfields}
 \end{figure}

The behavior of nematic liquid crystals inside spherical shells has been studied extensively for systems where the size of the particles is much smaller than that of the droplet, as is typical when molecular liquid crystals are used. In this limit, the spherical geometry of the droplet necessitates the formation of defects in the nematic director field of the liquid crystal. It has been shown via both theory and experiment that the favored defect structure for a nematic shell typically consists of four defects with topological charge $+1/2$. This defect structure leads to a baseball-like geometry of the nematic director field (see Fig. \ref{fig:directorfields}a). For spherical nematic shells, it has been shown that the four defects often lie on the vertices of a tetrahedron \cite{lubensky1992orientational,nelson2002toward}. However, their exact position is dependent on the elastic properties of the nematic phase, and for (nearly) hard rods at high density\cite{shin2008topological,bates2008nematic,dhakal2012nematic}, as well as for shells of smectic liquid crystals \cite{blanc2001confinement,lopez2012smectic}, it has been observed that these defects instead lie on a great circle. Another commonly observed defect structure is the generation of two defects of topological charge $+1$ at opposite poles of the shell, with the director field aligned along the pole-to-pole direction (see Fig. \ref{fig:directorfields}b). Additionally, both of these defect structures can in principle be inverted by rotating all particle directions by 90 degrees, resulting in an anti-baseball (Fig. \ref{fig:directorfields}c) and an anti-polar (Fig. \ref{fig:directorfields}d) geometry, respectively. While other structures have been observed in nematic shells~\cite{lopez2011frustrated,dzubiella2000topological}, these are typically less stable than the two- or four-defect geometries.

When considering colloidosomes, the size of the particles is typically small, but not negligible, when compared to the droplet size. As a result, the defect structure is strongly dependent on the ratio between droplet size and rod dimensions, and therefore on both the aspect ratio of the rods and the number of rods that form the shell. Here, we investigate close-packed structures of monodisperse short hard rods on a spherical surface. We map out the structures observed both in close-packed small clusters containing only a handful of particles, and in larger shells containing up a thousand particles. In order to obtain densely-packed structures, we use numerical optimization schemes, both with and without explicit biasing fields towards the typical defect structures known to occur for nematic shells. We find a rich variety of cluster structures depending on the number of rods and their aspect ratio $L/D$, where $L$ is the cylinder length and $D$ the diameter of the rods. For small clusters of up to around 100 rods, we find a sequence of cluster structures of varying symmetry, while for larger clusters we observe the formation of aligned domains of particles which are globally disordered for short rods ($L/D = 0.25$), but typically show a baseball-like global ordering for longer rods ($L/D = 0.5$ or $1$). Our results provide a route towards the creation of new colloidosomes out of rod-shaped particles, with both positional and orientational internal ordering.

The remainder of this paper is structured as follows. In Section \ref{sec:methods} we discuss the computational methods used for obtaining the densely packed clusters. In Section \ref{sec:small}, we describe the best packings obtained for clusters of up to 12 particles, while in Section \ref{sec:large} we report on the structure of larger clusters. Finally, in Section \ref{sec:conclusions} we discuss our conclusions.

\section{Methods} \label{sec:methods}

\begin{figure}
  \includegraphics[width=0.95\linewidth]{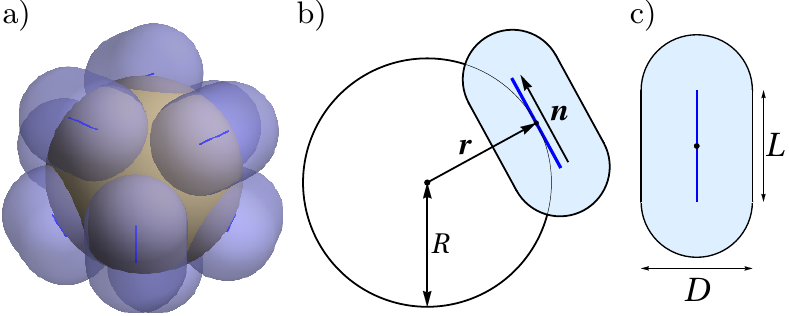}
  \caption{{\bf a)} Depiction of a configuration of 9 rods ($L/D = 0.5$) tangential to a central sphere. {\bf b)} Two-dimensional schematic of the model. Shown are the central sphere with radius $R$, a particles at position $\mathbf{r}$, and its orientation vector $\mathbf{n}$. {\bf c)} Length scales associated with a single rod: the cylinder length $L$ and the rod diameter $D$. }\label{fig:schematic}
\end{figure}

We investigate systems of $N$ hard spherocylinders with diameter $D$ and cylinder length $L$. For each rod, the position of its center of mass $\mathbf{r}$ is constrained to the surface of a sphere with radius $R$, and its direction vector ${\mathbf{n}}$ is confined to lie tangential to the surface of the sphere (see Fig. \ref{fig:schematic}). In this work, we focus on rods with an aspect ratio $L/D$ between 0 and 1, i.e. from spheres ($L/D = 0$) to rods with an aspect ratio of $L/D = 1$. In order to obtain the close-packed structure for different values of $N$, we have to find the minimum value of $R$ which allows a configuration without any overlapping rods. Such a constrained optimization problem can in principle be handled with a variety of techniques, including numerical optimization on a smoothened energy landscape which replaces the hard constraints~\cite{kottwitz1991densest,assoud2011penalty,ouguz2012packing}, simulated annealing \cite{krauth,uche2004concerning,shokef2009stripes}, and molecular dynamics simulations in which the particles slowly grow \cite{lubachevsky1990geometric,kansal2002computer}. 
Here, we use two approaches: a simulated annealing technique based on $NPT$ Monte Carlo (MC) simulations, performed at fixed number of rods $N$, pressure $P$, and temperature $T$, and the simulated compression of clusters using event-driven molecular dynamics (EDMD). 

In the MC approach, we simulate $N$ rods on a spherical shell with variable radius $R$ at constant temperature $T$, and use standard Monte Carlo volume moves \cite{bookfrenkel} to change the volume of the sphere, and therefore the radius. Additionally, we perform particle moves, where both the position $\mathbf{r}_i$ and orientation $\mathbf{n}_i$ of a particle are rotated randomly around the center of the sphere, and rotation moves, where a particle $i$ is rotated around the axis connecting the center of the sphere and the particle position $\mathbf{r}_i$  Overlaps are detected using the overlap algorithm introduced by Vega {\it et al.}\cite{vega1994fast}. The pressure associated with the acceptance of the volume moves is slowly increased during the simulation, up to a maximum of approximately $P=20k_B T/D^3$, with $k_B$ Boltzmann's constant. At this pressure, significant particle rearrangements are unlikely. After the maximum pressure has been reached, additional MC cycles are performed where volume moves are only allowed to decrease $R$, and never increase it, effectively corresponding to an infinite pressure. During the entire simulation, the step sizes associated with particle moves and rotations are continuously updated to maintain a reasonable acceptance ratio for these moves (around 30\%). In order to improve packing in large clusters, we perform these simulations both with and without an external biasing field, as will be described later in this paper.

In all cases, we allow the MC simulations to run for at least $\mathcal{O}(10^8)$ cycles, where each cycle consists of $N$ particle moves, $N$ particle rotations, and 2 volume moves. For each set of parameters ($N$, $L/D$), we perform five to ten independent runs, and select the final configuration with the highest density. For large clusters, longer simulation times tend to produce slightly better packings, and hence we do not claim that the packings obtained are strictly optimal for all investigated parameters. 

For the EDMD approach, we simulate approximate dynamics for hard rods confined to a spherical surface. In order to simplify and speed up the simulation, each rod is treated as a symmetric top rotating around a point fixed at the center of the sphere. Collisions between particles are predicted numerically \cite{de2007discontinuous}. Starting from a random initial configuration at low density, we decrease the radius $R$ of the sphere linearly with time at a constant speed $v_R$. Due to this compression, collisions between rods do not conserve energy, and as a result, the system is expected to heat up during the simulation, increasing the kinetic energy per particle. This reduces the numerical stability of the simulation. To counteract heating, we attempt to resolve each collision first in a way which conserves energy by neglecting the contribution of the shrinking surface to the relative velocity of the two particles. If this inelastic collision does not cause the particles to move apart after the collision, we resolve the collision normally instead. This significantly reduces the rate at which the kinetic energy increases. Additionally, we include an Andersen thermostat in our simulations: periodically, a random selection of particles are given a new angular momentum drawn from a Maxwell-Boltzmann distribution. 

The EDMD simulations are allowed to run until collisions can no longer be accurately predicted, which typically occurs when particles are extremely close-packed. In particular, we correct small numerical errors (i.e. overlaps) by restarting the simulation from a backup configuration using a smaller numerical gridsize for collision prediction \cite{de2007discontinuous}. If a more than ten of these corrections are required within one time unit, the simulation is stopped and the last non-overlapping configuration is taken as the close-packed configuration. As in the case of MC-based optimization, we typically perform five to ten independent runs to obtain a better estimate of the close-packed structure, with additional runs in regions where large differences between different runs are still observed, or when the packing density as a function of aspect ratio shows strong fluctuations.

Given equal amounts of CPU time, the EDMD-based packing optimization tended to slightly outperform the MC-based optimization on average, but not by a significant margin. It is likely that tuning parameters of both approaches can shift the balance between the two approaches. Note that for the biased systems, only MC-optimization was used, as extending the EDMD simulations to include (continuous) biasing potentials is not straightforward. 

The quality of a packing of rods is determined via the number density $\rho$, given by
\begin{equation}
 \rho = \frac{N}{4 \pi R^2}
\end{equation}
or, equivalently, via the surface coverage fraction $\phi$, defined as
\begin{equation}
 \phi = \rho A^\mathrm{rod}(R, L, D).
\end{equation}
Here, $A^\mathrm{rod}(R, L, D)$ is the surface area of the portion of the central sphere which is covered by a single rod. It is dependent on both the radius of the sphere and the rod aspect ratio, and is calculated via numerical integration. Note that in a flat plane, or equivalently when $R\to\infty$, the surface area covered by a rod is given by:
\begin{equation}
 A^\mathrm{rod}(R \to \infty, L, D) = D^2 \pi /4 + L D,
\end{equation}
and the maximum packing reachable for rods with a given $L/D$ is 
\begin{equation}
 \phi_{\mathrm{max}} = \frac{L/D + \pi/4 }{L/D + \sqrt{3}/2}.
\end{equation}

To test the reliability of our method, we also apply our optimization scheme to hard spheres (i.e. rods with $L/D = 0$). For cluster sizes up to $N=130$, the density of the best packed structures are known \cite{Sloane}, and can be used to test our algorithms. With the typical compression speeds and number of MC cycles we also use for rods, and the same amount of independent runs per cluster size, we recover 90\% of the optimal packings for clusters up to size $N=130$ within a density difference of $0.001 D^{-2}$, with most of the larger deviations occuring for large $N$. Moreover, for the majority of choices of $N\le130$, we recover the best known packing within $10^{-6} D^{-2}$. For aspect ratios $L/D >0$, the number of degrees of freedom to optimize is a factor $3/2$ higher than in the case of spheres, and as a result, we do not expect to find optimal packings for large $N$. Nonetheless, we expect to reliably find the best-packed structure for most clusters up to sizes of around $N\simeq50$, and good approximations of the best packing for larger numbers of rods.

\section{Small clusters ($N \le 12$)}\label{sec:small}

\newcolumntype{V}{>{\centering  \arraybackslash \vspace{2pt}} m{0.08 \linewidth} }
\newcolumntype{L}{>{\raggedleft \arraybackslash \vspace{2pt}} m{0.02 \linewidth} }
\newcolumntype{K}{>{\centering \arraybackslash \vspace{2pt}} m{0.08 \linewidth} }
\newcolumntype{X}{>{\raggedright \arraybackslash \vspace{2pt}} m{10cm} }
\newlength{\PS}
\setlength{\PS}{0.81cm}
\begin{table*}
\begin{center}
  \includegraphics[width=0.95\linewidth]{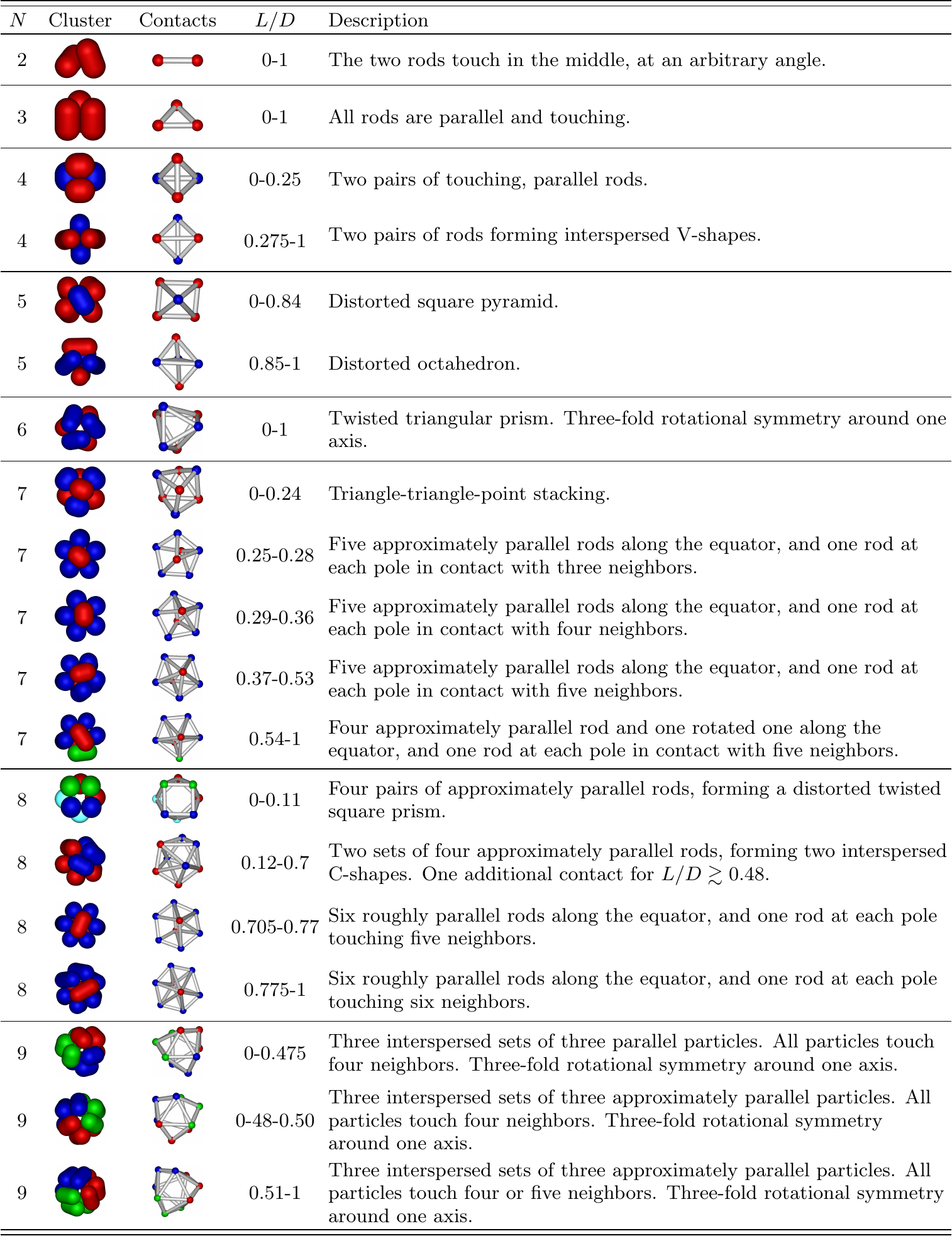}
\end{center}
\caption{Different cluster types with the best obtained packing for small clusters of rods with aspect ratio $L/D$ between 1 and 2. The columns list, in order, the number of particles $N$, a snapshot of a typical cluster, a schematic of the positions of the center of mass of the rods with lines connecting touching rods (surface-to-surface cutoff radius $r_c = 0.0025~D$), the values of $L/D$ where this structure was observed to be the best packing, and a description of the cluster. Colors in the snapshots serve as a guide to the eye.} \label{pictable1}
\end{table*}

\begin{table*}[ht]
\begin{center}
  \includegraphics[width=0.95\linewidth]{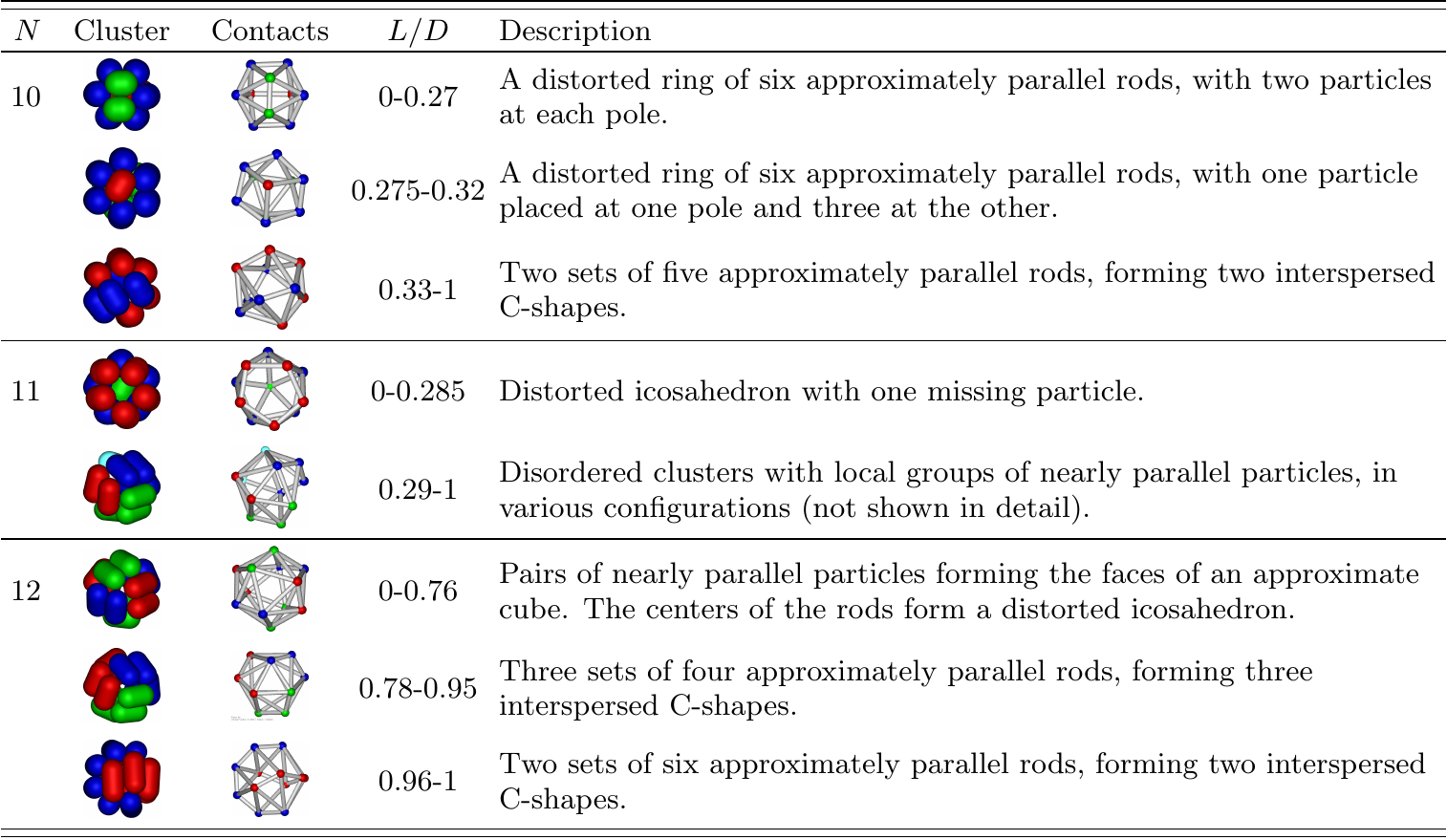}
\end{center}
\caption{Different cluster types with the best obtained packing for small clusters of rods with aspect ratio $L/D$ between 1 and 2. Note that for $N=10$ and $11$, most clusters show little symmetry, and the ordering or number of contacts between particles in the cluster can change with small changes in $L/D$. } \label{pictable2}
\end{table*}

For small clusters up to size $N=12$, we extensively explored the optimal packing configurations over the aspect ratio range $L/D = 0$ to 1, typically in steps of $0.01$. For each value of $N$, we performed both EDMD and MC based packing optimizations. In addition, in regions where several structures compete, we perform optimizations starting from configurations taken from optimal packings at a slightly higher or lower aspect ratio. For most cluster sizes, we find several distinct structures which maximize the packing, as a function of the aspect ratio. In Tables \ref{pictable1} and \ref{pictable2}  we report the best packing structures for all investigated cluster sizes, and include snapshots as well as schematic illustrations of the contact graph for each described structure, where we plot the particles as small spheres with lines connecting touching particles. To determine whether two rods are in contact, we here apply a simple cutoff criterium, where contact implies that the surface-to-surface distance between the two rods is less than $r_c = 0.0025D$. We classified the different structures based on the sets of contacts, sets of parallel rods, and visual inspection.

As can be seen in Tables \ref{pictable1} and \ref{pictable2}, we observe a rich variety of different structures even for a small number of particles, resulting from the interplay between the shape of the particles and the spherical geometry. This is consistent with the variety of symmetries in the packings observed in the disks or spheres on spherical surfaces \cite{kottwitz1991densest,Sloane}, and in spheres \cite{de2015entropy} or anisotropic particles \cite{peng2013colloidal,teich2016clusters} confined to the inside of a sphere. Of particular interest are the rotationally symmetric structures observed for $N=6$, $N=9$, and $N=12$. In each case, we find chiral structures with three-fold symmetry for a wide range of aspect ratios, reminiscent of experimental realizations of close-packed clusters of dumbbell-shaped particles \cite{peng2013colloidal}. Additionally, we observe baseball-like director field geometries for $N=8$, $10$, and $12$, where each domain is made up of a single row of aligned particles. Note that for $N=10$ and $11$, the structure of the optimally packed configuration is highly dependent on the exact aspect ratio, and in many cases several different structures compete closely for optimum packing. Hence, we only report the general structure of the observed clusters and a typical depiction of the contacts, rather than list each variation exhaustively.

\begin{figure}
 \includegraphics[width=\linewidth]{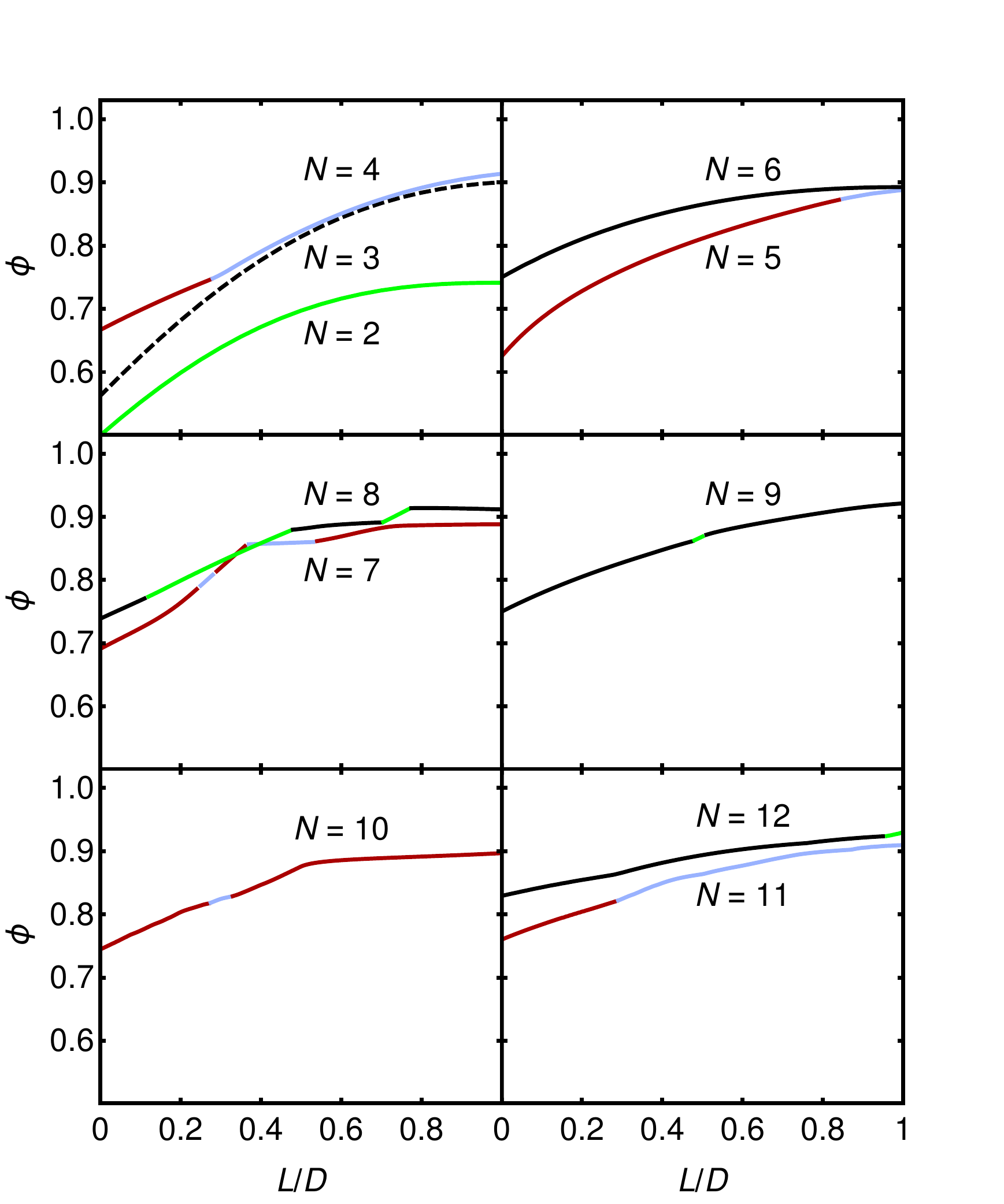}
 \caption{Maximum obtained area packing fraction $\phi$ for small clusters as a function of the aspect ratio $L/D$ for different cluster sizes as indicated. Different colors along the same line indicate different structures as indicated in Tables \ref{pictable1} and \ref{pictable2}. }\label{fig:smallcluspacking}
\end{figure}

In Fig. \ref{fig:smallcluspacking} we plot the best obtained surface area fraction $\phi$ as a function of the aspect ratio $L/D$ for the investigated cluster sizes. Generally, the surface packing fraction of the cluster increases with increasing aspect ratio, with the only negative slope observed for $N=8$ at high aspect ratios. In most cases, the packing increases smoothly, with sharp cusps at transitions from one structure to another, as indicated by the changes in color, corresponding to the regimes described in Tables \ref{pictable1} and \ref{pictable2}. For $N=10$ and $N=11$, the packing curves are less smooth, owing to changes in cluster geometry not included in the Tables. For all investigated aspect ratios, clusters of 12 particles exhibit the highest surface packing ratio.

\section{Large clusters}\label{sec:large}

\newcommand{\zhat}{\hat{\mathbf{z}}}
\newcommand{\yhat}{\hat{\mathbf{y}}}
\newcommand{\rhat}{\hat{\mathbf{r}}}
\newcommand{\nhat}{\hat{\mathbf{n}}}
\newcommand{\qpol}{q_\mathrm{p}}
\newcommand{\qbb}{q_\mathrm{bb}}
\newcommand{\qloc}{q_\mathrm{loc}}

We now turn our attention to larger clusters, with $L/D$ ranging up to $N=1000$ particles. In order to limit parameter space, we here restrict ourselves to four aspect ratios $L/D=0$ (i.e. spheres), $0.25$, $0.50$ and $1$. For each aspect ratio, we investigate a selection of cluster sizes using unbiased EDMD and MC compression approaches as well as biased MC compression. In the biased approach, we enhance the MC approach described earlier with an additional biasing potential, which aligns the rods with a chosen orientation field. The biasing potential is given by
\begin{equation}
 V^\mathrm{ext} = -\epsilon \sum_{i=1}^N \left| \hat{\mathbf{n}}_i \cdot \hat{\mathbf{b}}(\mathbf{r}_i)\right|,\label{eq:biaspot1}
\end{equation}
where $\mathbf{r}_i$ and $\hat{\mathbf{n}}_i$ are vectors corresponding to the location and axial direction of particle $i$, respectively, and $\hat{\mathbf{b}}(\mathbf{r})$ is the (location-dependent) biasing direction. Additionally, $\epsilon$ is the strength of the biasing potential. For a polar orientation field, we simply set
\begin{equation}
 \hat{\mathbf{b}}_\mathrm{p}(\mathbf{r}) = \zhat_\parallel(\mathbf{r}),\label{eq:polarfield}
\end{equation}
where $\zhat_\parallel(\mathbf{r})$ is a unit vector indicating the projection of the $\zhat$ axis onto the sphere surface at position $\mathbf{r}$, i.e.
\begin{equation}
\zhat_\parallel(\mathbf{r}) = \frac{ \zhat - (\zhat \cdot \rhat) \rhat}{\sqrt{1-(\zhat\cdot\rhat)^2}},
\end{equation}
with $\rhat$ a unit vector in the direction of $\mathbf{r}$.
Similarly, for a baseball geometry, we choose
\begin{equation}
 \hat{\mathbf{b}}_\mathrm{bb}(\mathbf{r}) = \left\{ \begin{array}{cc}
                                        \zhat_\parallel(\mathbf{r}) & \mathrm{if} \,\, x < 0 \\
                                        \yhat_\parallel(\mathbf{r}) & \mathrm{if} \,\, x > 0 \\
                                        \end{array} \right.. \label{eq:bbfield}
\end{equation}
Here, $x$ is the $x$-coordinate of the particle under consideration, where we have assumed that the central sphere is centered at the origin. In addition, we have performed optimizations simply using $\zhat$ and $\yhat$ instead of $\zhat_\parallel$ and $\yhat_\parallel$, but not see qualitatively different results.

For each chosen value of $N$ and $L/D$, we vary the strength of the biasing potential between $\epsilon = 1$ and $5k_BT$ in steps of $1k_B T$ in the MC simulation, and  select the highest density structure from multiple runs at each field strength. Typically, the best results are obtained for $\epsilon = 2-3k_BT$, although this varies between state points. In addition, we have investigated the effect of biasing the rods to lie perpendicular to the described orientation field by setting $\epsilon$ between $-1$ and $-5k_BT$, resulting in anti-polar and anti-baseball geometries. However, the densities in these structures were always lower than those obtained for positive values of $\epsilon$, and hence we omit these results here.

\begin{figure}
 \includegraphics[width=\linewidth]{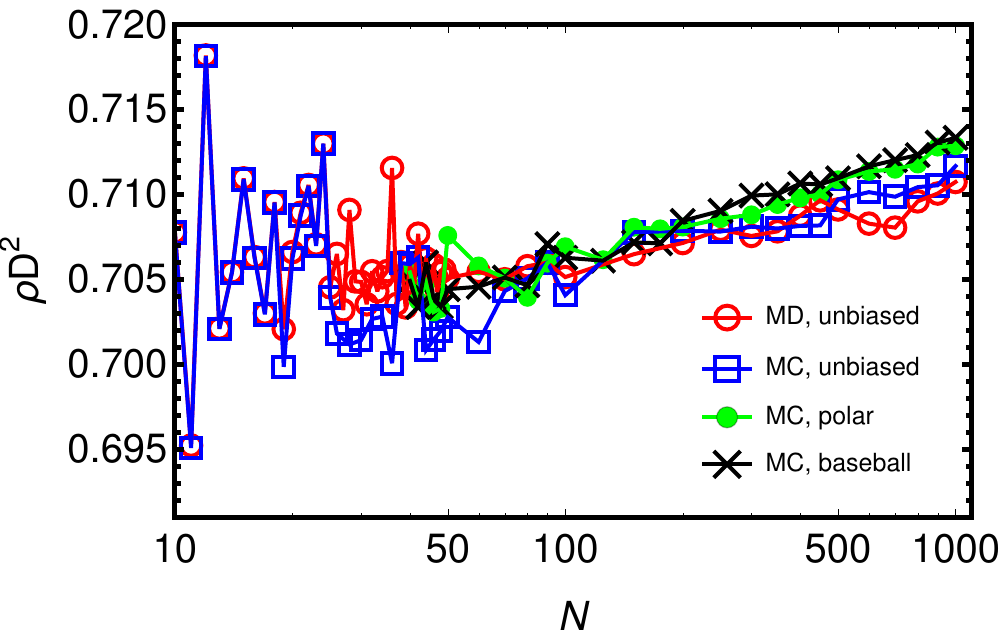}
 \caption{Maximum obtained surface density $\rho$  as a function of the cluster size $N$ for aspect ratio $L/D = 0.5$. The lines indicate different optimization methods as indicated, with points specifying all investigated choices of $N$.}\label{fig:largepacking0p5}
 \end{figure}

In Fig. \ref{fig:largepacking0p5}, we plot the highest density per cluster size obtained via four different methods: unbiased EDMD, unbiased MC, biased MC with polar geometry, and biased MC with baseball geometry. For small clusters (up to $N \simeq 50$), our unbiased approaches (red circles and blue squares) provide the best results, while for larger clusters we typically obtain the best packed structures in the biased simulations. For smaller clusters, the maximum surface density is highly dependent on $N$, as evidenced by the sharp fluctuations in this regime. At certain ``magic numbers'' of particles, the packing shows sharp peaks, typically corresponding to clusters of high symmetry. As the number of particles per cluster increases, small changes in the number of particles have less effect on the overall cluster structure and packing, resulting in smoother behavior of the maximum density as a function of $N$.  While for a small number of intermediate cluster sizes the optimal packing was obtained via a polar biasing field (green points), for sufficiently large clusters the optimal obtained packing consistently corresponds to a baseball geometry (black squares).

\begin{figure}
 \includegraphics[width=\linewidth]{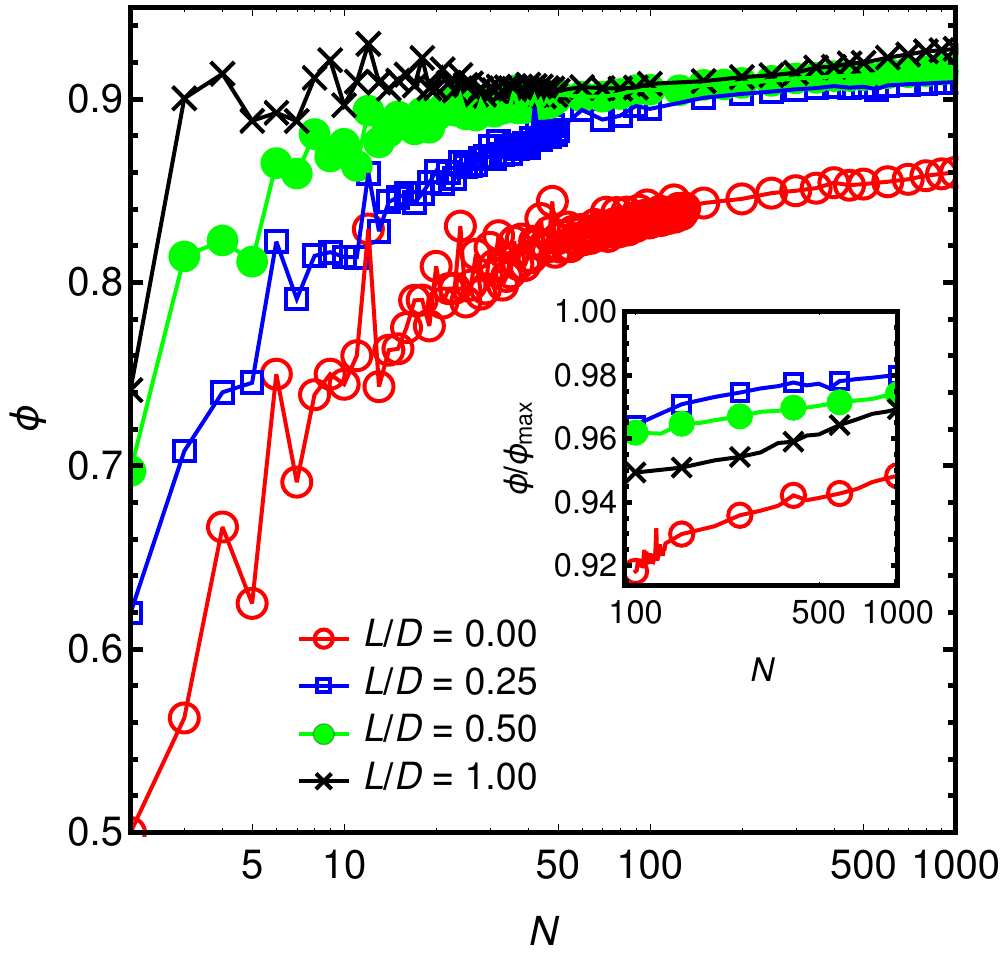}
 \caption{Maximum obtained surface area coverage fraction $\phi$ as a function of the cluster size $N$ for different aspect ratios $L/D$ as indicated. The inset shows the same data normalized by the maximum packing fraction $\phi_\mathrm{max}$ attainable for each aspect ratio.}\label{fig:largecluspacking}
\end{figure}

In Fig. \ref{fig:largecluspacking}, we plot the maximum obtained surface packing fraction $\phi$ for four different aspect ratios: $L/D = 0$ (spheres), $0.25$, $0.5$, and $1$. For all investigated cluster sizes, we observe that longer rods can reach a higher maximum packing. However, this effect is mainly caused by the fact that longer rods can reach higher packing fractions even in a planar geometry. As shown in the inset, when we compare the surface packing of the rods to the maximum packing obtainable in a flat plane ($\phi_\mathrm{max}$), we see that short rods ($L/D = 0.25$) consistently reach surface fractions closer to $\phi_\mathrm{max}$ than both spheres and longer rods. Thus, although the curved substrate decreases the maximum packing for all aspect ratios, the effect is minimized for short rods, suggesting that around $L/D = 0.25$, the rods are better capable of incorporating the defects dictated by the geometry into a close-packed structure due to their additional degrees of freedom. This is reminiscent of the observation that the {\em random} close packing of ellipsoidal particles shows a maximum at low aspect ratios \cite{donev2004improving,man2005experiments,sacanna2007observation}.

We now turn our attention to the structures of the best packed clusters. In order to quantify the global ordering of the cluster into either a polar or baseball geometry, we use two order parameters ($\qpol$ and $\qbb$, respectively) based on the energy described in Eq. \ref{eq:biaspot1}. In particular, we define for the polar order parameter:
\begin{eqnarray}
 \qpol = \frac{-V^\mathrm{ext}_p / N \epsilon - 2/\pi}{1 - 2/\pi}.
\end{eqnarray}
Here, $V^\mathrm{ext}_\mathrm{p}$ is the potential energy in Eq. \ref{eq:biaspot1} for the polar orientation field in Eq. \ref{eq:polarfield}, minimized with respect to the global rotations of the cluster. For the baseball geometry, we define $\qbb$ in an identical fashion. The offset and normalization in the definition of the order parameter is chosen such that for a fully disordered system, such as an ideal gas of rods confined to the spherical surface, $\qpol = \qbb = 0$ for sufficiently large $N$, while for a set of particles perfectly aligned with the chosen orientation field, $\qpol = \qbb = 1$. 

In addition, we define a local aligning parameter $\qloc$ as
\begin{equation}
 \qloc = \frac{1}{(1-2/\pi)N} \sum_{i=1}^N \frac{1}{N_b(i)} \sum_{j \in \{N_b(i)\}} \left| \nhat_i \cdot \nhat_j \right| - 2/\pi, \label{eq:localalignment}
\end{equation}
where $\{N_b(i)\}$ is the set of nearest neighbors of particle $i$, defined as all particles with a surface-to-surface distance to $i$ which is less than $0.5~D$. This order parameter measures the degree of local alignment within the cluster, and is equal to 1 in a system of perfectly aligned particles, and is zero in a disordered gas in a flat plane. As might be expected, the overall ordering is typically higher in the systems composed of longer rods (higher $L/D$), in particular for large clusters. This can be understood from the consideration that the free volume gained from aligning longer rods is higher than for shorter ones, favoring larger aligned domains with fewer defects.

\begin{figure}
 \includegraphics[width=\linewidth]{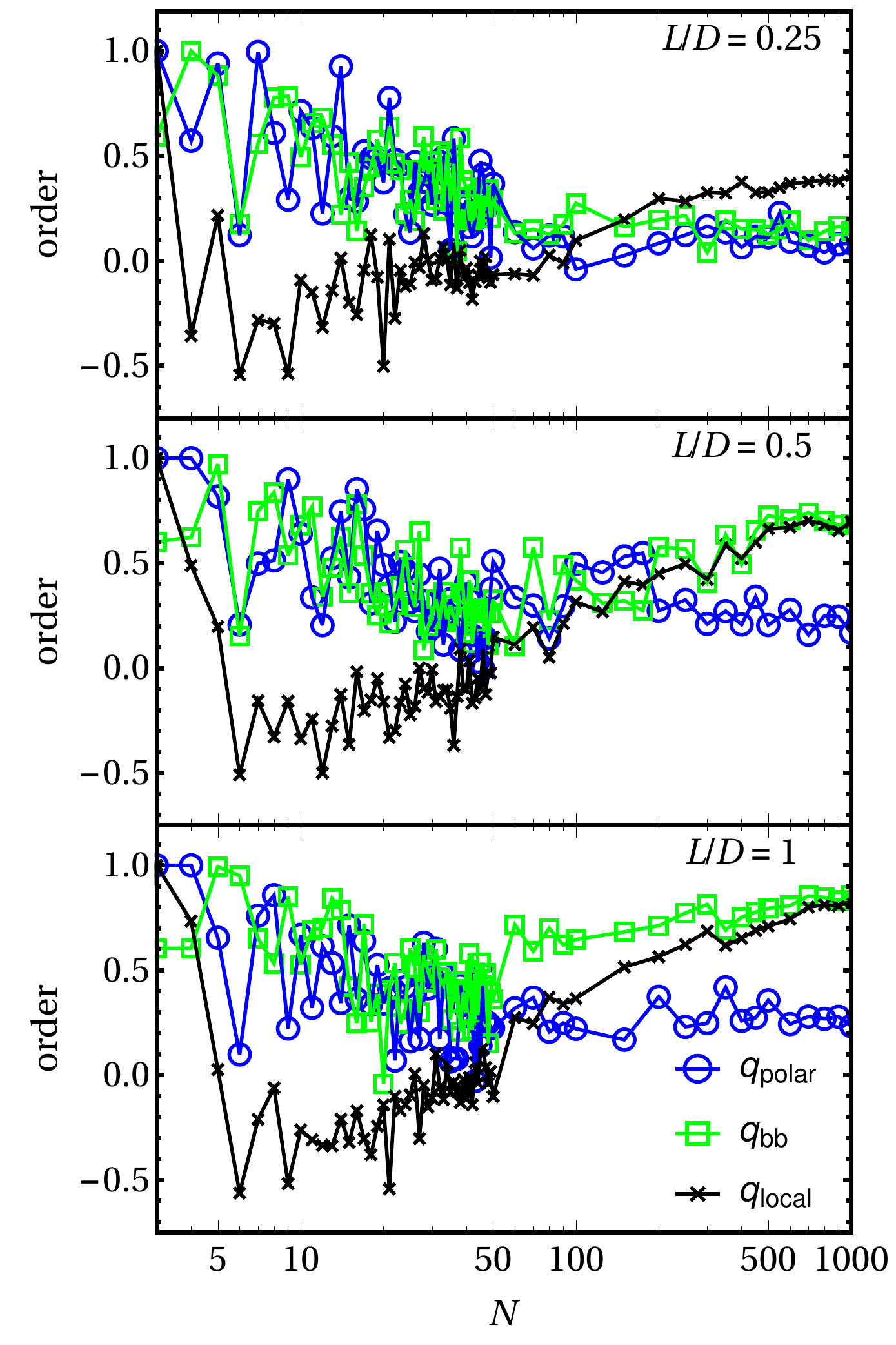}
 \caption{Order parameters associated with polar geometry ($\qpol$), baseball geometry ($\qbb$) and local alignment ($\qloc$) as a function of cluster size for the rod clusters with the maximum obtained packing. }\label{fig:orderparas}
\end{figure}

In Fig. \ref{fig:orderparas}, we plot these order parameters for the densest obtained cluster at all investigated combinations of aspect ratio and cluster size. Note that we have omitted data for $L/D = 0$, as the described order parameters are only meaningful for rods. We observe different ordering scenarios for each of the three investigated aspect ratios. 

For short rods ($L/D = 0.25$), only small clusters show significant global order, which is strongly dependent on $N$. For larger clusters both $\qpol$ and $\qbb$ approach zero, indicating a lack of global ordering in the cluster. Note that in the latter regime, biasing the system during packing optimization provided effectively no benefit. In contrast, the local alignment, as expressed by $\qloc$ is close to zero or negative for small clusters, but grows significantly as a function of cluster size. 
Slightly longer rods, i.e. $L/D = 0.5$, exhibit a similar strongly fluctuating degree of global ordering at small $N$, but favor configurations with polar geometry for several cluster sizes around $N=100$. For larger $N$, baseball geometry is clearly favored, and $\qbb$ gradually increases while $\qpol$ decreases. Again, the degree of local alignment $\qloc$ increases as a function of $N$, and for high $N$ shows significant correlation with the global order parameter $\qbb$. 
Finally, for rods with $L/D = 1.0$, we no longer observe a region where polar order dominates.

\begin{table*}[ht]
\begin{center}
  \includegraphics[width=\linewidth]{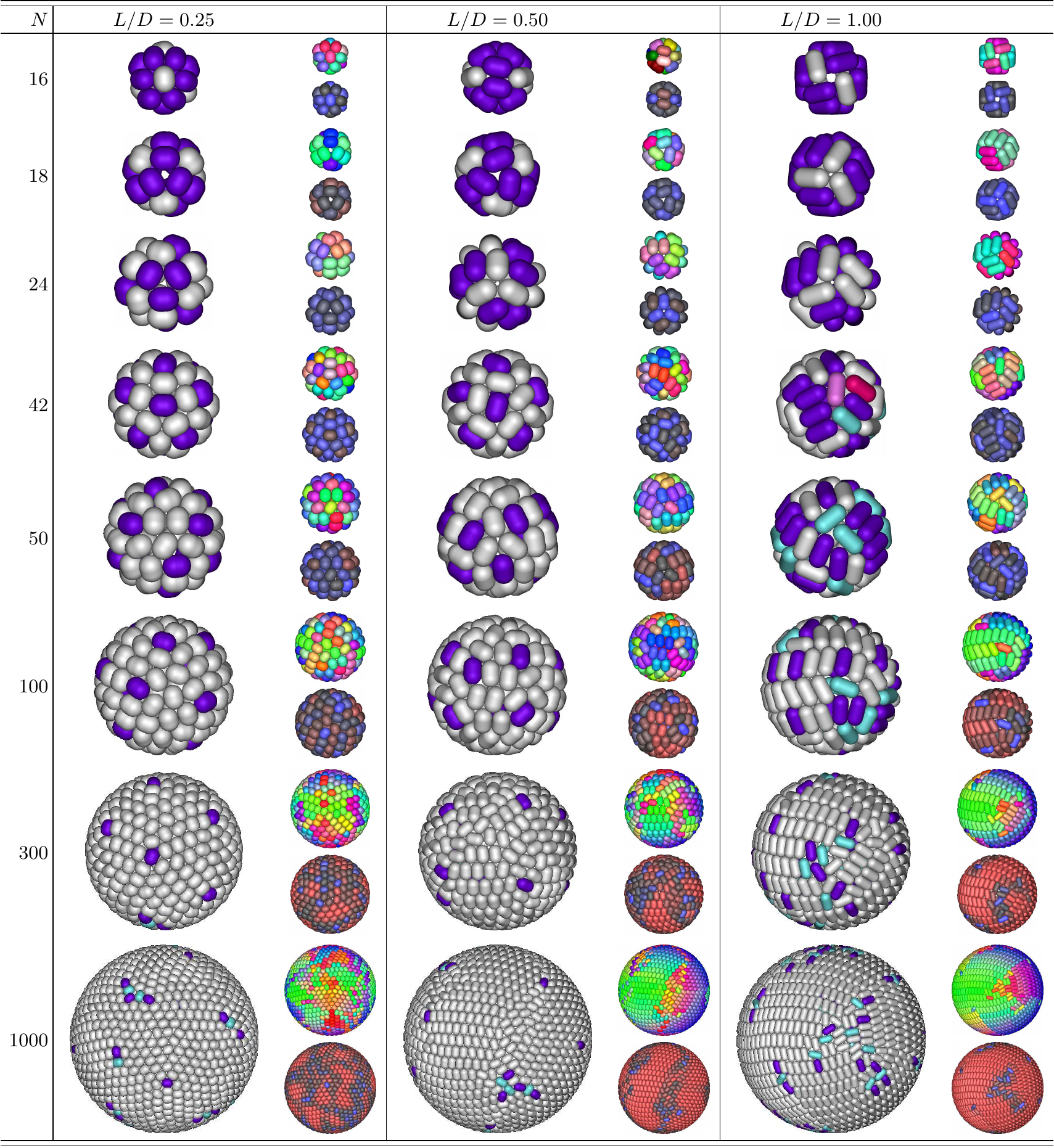}
\end{center}
\caption{Best obtained packings for a range of cluster sizes. For each cluster size $N$ and aspect ratio $L/D$, we show three pictures of the cluster from the same viewpoint. The larger image on the left shows the particles colored based on the number of neighbors. The smaller images show the particles colored based on their individual orientation (top right) and the their contribution to the local aligning order parameter $\qloc$ (bottom right), with red indicating strong local aligning and blue weak alignment. In case of a baseball geometry, the clusters are viewed directly from the direction of one of the topological defects. Additionally, the cluster for $L/D = 0.50$ and $N=100$ is rotated such that its polar ordering is along the vertical axis.} \label{bigclustable}
\end{table*}

In order to illustrate the types of packings observed in these larger clusters, we show in Table \ref{bigclustable} snapshots of clusters for a range of cluster sizes at each investigated aspect ratio. To showcase the various types of ordering, we show each snapshot in three different colorings. In the larger images, we color each particle based on the number of nearest neighbors it has, as determined from a Voronoi construction on the centers of mass of all rods. In the closest-packed configuration in a flat plane, each particle would have six nearest neighbors (light gray particles in Table \ref{bigclustable}). However, the spherical geometry necessitates the creation of at least 12. For $L/D=0.25$ we indeed observe exactly 12 particles with five neighbors each (dark blue in in Table \ref{bigclustable}) in the majority of the best packed clusters up to size $N=100$ ($>80\%$ of the investigated sizes), typically arranged roughly on the vertices of an icosahedron. For larger clusters, these point defects gradually transform into grain boundaries consisting of lines of particles alternating between five and seven (light blue) neighbors, strongly reminiscent of the known behavior for shells of spherical particles \cite{bausch2003grain}. For $L/D = 0.5$, we observe similar behavior, although the point defects are generally distributed more unevenly over the surface of the sphere. At aspect ratio $L/D=1.00$, we observe significantly more defects in the Voronoi structure for small clusters, and larger domains of aligned rods in the larger clusters. As is clear from Table \ref{bigclustable}, a number of the clusters with $N \lesssim 60$ show high degrees of symmetry, especially for the smaller aspect ratios. 

In addition to the Voronoi coloring, we show in Table \ref{bigclustable} snapshots of the cluster based on the individual orientation of each particle (top right image for each cluster), as well as an image colored based on the local degree of alignment as calculated via the contribution of each particle to Eq. \ref{eq:localalignment}. As the images show, the degree of alignment increases with both cluster size and rod length, as might be expected. For $L/D = 0.25$, only small domains are observed even in the largest clusters, and these domains show no noticeable global ordering. For $L/D = 0.50$ and $N=100$, the polar ordering (along the vertical axis in the image) can be seen, but is clearly weak. In contrast, the baseball geometry in the larger clusters for both $L/D = 0.50$ and $1.00$ are clearly visible, and are accompanied by large ordered domains.

\section{Conclusions}\label{sec:conclusions}

In short, we have explored the close packing of rod-shaped particles confined to a spherical surface. We find a wide range of close-packing structures exists based on the number ($N$) and aspect ratio ($L/D$) of the rods. In particular, for small clusters we find a rich variety of (often symmetrical) structures which depend strongly on the exact values of $N$ and $L/D$. Larger clusters instead show global ordering dominated by domains of aligned particles, which for $L/D \ge 0.50$ is typically arranged in a baseball-like geometry, reminiscent of the behavior expected in nematic clusters \cite{shin2008topological,bates2008nematic,blanc2001confinement,lopez2012smectic}. In addition, for $L/D = 0.5$ we observe an intermediate regime where polar ordering is favored. Since the spherical geometry necessitates the formation of defects in the packing of the rods, the maximum packing of rods on a sphere is lower than it would be in a planar geometry. Interestingly, the effect of the curvature is smaller for short rods ($L/D \simeq 0.25$) than for either spheres or longer rods.

The packings found in this work can in principle be realized on the colloidal scale by e.g. trapping rod-shaped colloidal particles on the surface of evaporating emulsion droplets \cite{manoharan2006colloidal}. The use of rod-shaped particles instead of spheres for these colloidosomes provides access to a wider range of cluster structures, with distinct symmetries and pore distributions. Another possibility might be to confine macroscopic granulate particles between spherical shells, and to employ shaking or tapping to reach a close-packed state\cite{donev2004improving}. We note here that the densely packed baseball-geometry observed for longer rods does not reliably occur spontaneously in our simulations without the presence of external biasing fields. As a result, future work on this topic should address the behavior of the rods at finite pressures, where cage-breaking is still possible, to investigate when and how this geometry can be expected to arise spontaneously. Another future avenue for research is the packing of rods on manifolds with geometries other than a sphere, such as toroidal \cite{bowick2009two}, cylindrical \cite{mughal2012dense,mughal2014theory,napoli2012extrinsic}, ellipsoidal \cite{bates2010defects}, or hyperbolic surfaces \cite{irvine2010pleats}.

\section{Acknowledgements}
We gratefully acknowledge funding from the Alexander von Humboldt foundation, and from the  Deutsche Forschungsgemeinschaft, within project L0418/20-1.

\bibliography{rods}

\end{document}